\pdfoutput=1
\documentclass[aps,twocolumn, floatfix, prl]{revtex4}
\usepackage{graphicx}
\DeclareGraphicsExtensions{.pdf}
\usepackage{amsmath,amssymb,bbold,bm,color}
\usepackage{float}
\usepackage{epstopdf}

\newcommand{\bk}{{\bm k}}
\newcommand{\bK}{{\bm K}}
\newcommand{\bq}{{\bm q}}

\newcommand{\bp}{{\bm p}}

\newcommand{\bv}{{\bm v}}
\newcommand{\bu}{{\bm u}}

\newcommand{\bB}{{\bm B}}

\newcommand{\bb}{{\bm b}}
\newcommand{\bA}{{\bm A}}

\newcommand{\btau}{{\bm \tau}}

\newcommand{\cA}{{\cal A}}

\newcommand{\bee}{\begin{equation}}
\newcommand{\ee}{\end{equation}}

\begin{document}

\title{Quantum oscillations without magnetic field}

\author{Tianyu Liu}
\author{D. I. Pikulin}
\author{M. Franz}
\affiliation{Department of Physics and Astronomy, University of
British Columbia, Vancouver, BC, Canada V6T 1Z1}
\affiliation{Quantum Matter Institute, University of British Columbia, Vancouver BC, Canada V6T 1Z4}

\begin{abstract} 
When magnetic field $B$ is applied to a metal, nearly all observable
quantities exhibit oscillations periodic in $1/B$. Such quantum
oscillations reflect the fundamental reorganization of electron states into Landau levels as a canonical response
of the metal to the applied magnetic field. We predict here that,
remarkably, in the recently discovered Dirac and Weyl semimetals
quantum oscillations can occur in the complete absence of magnetic
field. These zero-field quantum oscillations are driven by elastic
strain which, in the space of the low-energy Dirac fermions, acts as a
chiral gauge potential. We propose an experimental setup in  which the strain in a
thin film (or nanowire) can generate pseudomagnetic field $b$
as large as 15T and demonstrate the resulting de Haas-van Alphen and 
Shubnikov-de Haas oscillations periodic in $1/b$.

\end{abstract}

\date{\today}

\maketitle

Dirac and Weyl semimetals \cite{Savrasov2011,burkov2011b,Vafek2014}
are known to exhibit a variety of exotic behaviors owing to their
unusual electronic structure comprised of linearly dispersing electron
bands at low energies. This includes the pronounced negative
magnetoresistance \cite{fukushima2008,son2013,
  kim2013,huang2015,ong2015,burkov2015,valla2016,jia2016}  attributed
to the phenomenon of the chiral anomaly
\cite{adler1969,bell1969,nielsen1983},  theoretically predicted
nonlocal transport \cite{parameswaran2014,baum2016}, Majorana flat bands
\cite{anffany2016}, as well as an unusual type of quantum oscillations (QO)
that involve both bulk and topologically protected surface states
\cite{Potter2014,Moll2016}.  In this theoretical study we establish a
completely new mechanism for QO in Dirac and Weyl
semimetals that requires no magnetic field. These zero-field
oscillations occur as a function of the applied elastic strain and,
similar to the canonical de Haas-van Alphen and Shubnikov-de Haas
oscillations \cite{shoenberg_book}, manifest themselves as
oscillations periodic in $1/b$, where $b$ is the strain-induced
pseudomagnetic field, in all measurable thermodynamic and transport
properties. To the best of our knowledge this is the first instance of such
zero-field quantum oscillations in any known substance.

\begin{figure}[t]
\includegraphics[width = 8.0cm]{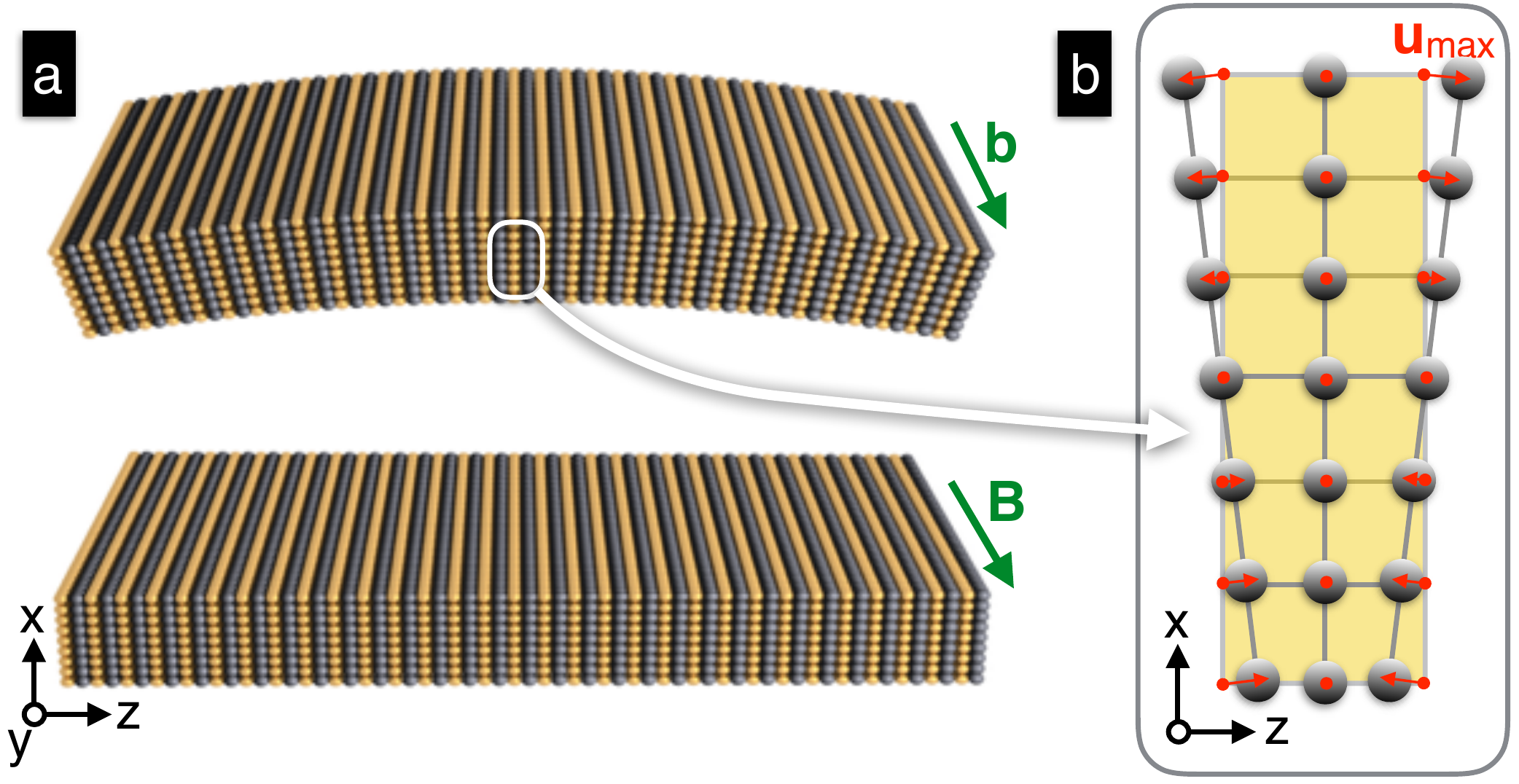}
\caption{Proposed setup for strain-induced quantum oscillation
  observation in Dirac and Weyl semimetals. a) Bent film is
  analogous, in terms of its low-energy properties, to an unstrained
  film subject to magnetic field $B$. b) Detail of the atomic
  displacements in the bent film. Displacements have been exaggerated
  for clarity.
}\label{fig1}
\end{figure}

Materials with linearly dispersing electrons respond in peculiar
ways to the externally imposed elastic strain.  In graphene, for
instance, the effect of curvature is famously analogous to a
pseudomagnetic field \cite{guinea2010} that can be quite large and is
known to generate pronounced Landau levels observed in the tunneling spectroscopy
\cite{levy2010}. Recent theoretical work
\cite{shapourian2015,cortijo2015,fujimoto2016,pikulin2016,Grushin2016} showed that similar
effects can be anticipated in three-dimensional Dirac and Weyl
semimetals, although the estimated field strengths in the geometries
that have been considered are rather small (below 1 Tesla in Ref.\
\cite{pikulin2016}). Ordinary quantum
oscillations, periodic in $1/B$, have already been observed in Dirac
semimetals Cd$_3$As$_2$ and Na$_3$Bi \cite{he2014,liu2015,ong2016,Moll2016} but the magnetic field required
is $B\gtrsim 2$T. This, then, would seem to rule out the observation
of strain-induced QO in geometries considered
previously. We make a key advance in this work by devising a
new  geometry in which pseudomagnetic field $b$ as large as 15T can
be achieved. The proposed setup consists of a thin
film (or a nanowire) in which pseudomagnetic field $b$ is generated by
a simple bend as illustrated in Fig.\ \ref{fig1}. 
%An important requirement is that the vector $\bK$  connecting two
%Dirac/Weyl points (Fig.\ \ref{fig0}) has a non-vanishing projection
%$\tilde{\bK}$ onto the plane of the film and the axis of the bend is perpendicular to $\tilde{\bK}$.
%

\begin{figure}[t]
\includegraphics[width = 6.5cm]{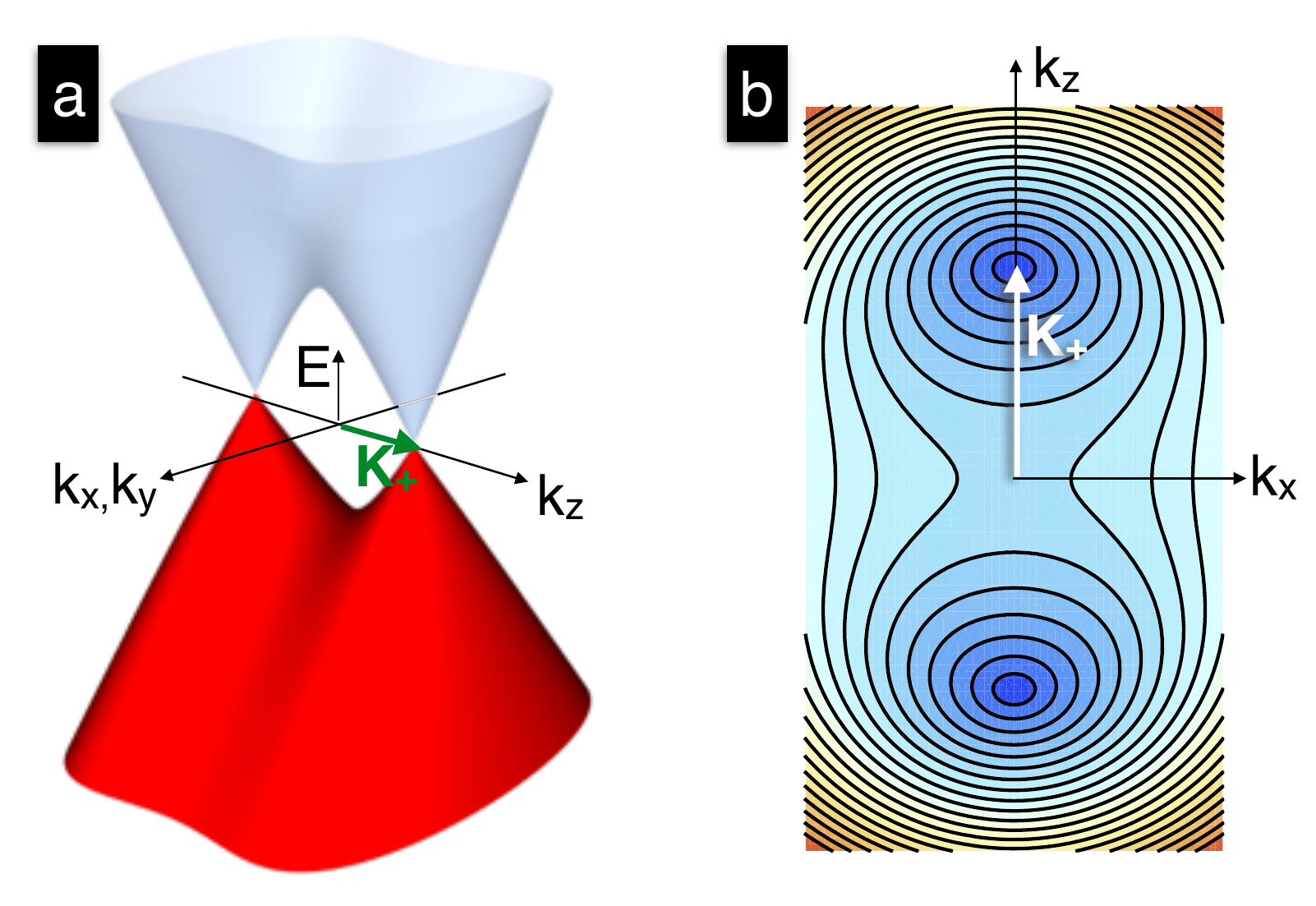}
\caption{Schematic depiction of
  the low-energy electron excitation  spectrum in Dirac and Weyl
  semimetals. a) In a Dirac semimetal the bands are doubly degenerate due
  to the spin degree of freedom while in a Weyl semimetal they are
  nondegenerate. b) Contours of constant energy for $k_y=0$. For
  magnetic field $\bB\parallel \hat{y}$ these correspond to the extremal
  orbits \cite{shoenberg_book} that give rise to QO periodic in $1/B$.
}\label{fig0}
\end{figure}

For simplicity and concreteness we focus in the following on Dirac
semimetal Cd$_3$As$_2$
\cite{zhizhun2013,borisenko2014,neupane2014,jeon2014,he2014,liu2014b}
which is the best characterized representative of this class of
materials. Our results are directly applicable also to Na$_3$Bi
\cite{zhizhun2012,yulin2014,yulin2014b} whose low-energy description
is identical, and are easily extended to other Dirac and Weyl
semimetals \cite{hasan2015,ding2015,yan2015,chen2015,xu2015}. We start from the tight-binding model formulated in Refs.\
\cite{zhizhun2013,zhizhun2012} which describes the
low-energy physics of Cd$_3$As$_2$  by including the band inversion of its atomic
Cd-$5s$ and As-$4p$ levels near the $\Gamma$ point. In the basis of
the spin-orbit coupled states $|P_{3\over 2},{3\over 2}\rangle$,
$|S_{1\over 2},{1\over 2}\rangle$, $|S_{1\over 2},-{1\over 2}\rangle$
and $|P_{3\over 2},-{3\over 2}\rangle$ the model is defined by a $4\times 4$
matrix Hamiltonian
\bee \label{h1}
 H^{\rm latt} =\epsilon_\bk+
\begin{pmatrix}
h^{\rm latt} & 0 \\
0 & -h^{\rm latt}
\end{pmatrix},
\ee
on a simple rectangular lattice with spacings $a_{x,y,z}$, where 
\bee \label{h3}
h^{\rm latt}(\bk)=m_\bk\tau^z+\Lambda(\tau^x\sin{a_x k_x}+\tau^y\sin{a_y k_y}),
\ee
$\btau$ are Pauli matrices in the orbital space and 
$m_\bk=t_0+t_1\cos{a_z k_z} +t_2(\cos{a_x k_x}+\cos{a_y k_y})$. For the analytic calculations below we will assume $a_i=a$, while in numerics we will use the actual lattice constants of Cd$_3$As$_2$.
Various tunneling amplitudes and $\epsilon_\bk$ are given in
Supplementary Material (SM). The low-energy spectrum of $h^{\rm latt}$ consists of a pair
of Weyl points, shown in Fig.\ \ref{fig0}a, which carry opposite chirality $\eta=\pm 1$
and are located at  crystal momenta $\bK_\eta=(0,0,\eta Q)$ with $Q$ given by
$\cos(aQ)=-(t_0+2t_2)/t_1$.  The lower diagonal block in Eq.\
(\ref{h1}) describes the spin-down sector in Cd$_3$As$_2$ and has
identical spectrum. Terms in $\epsilon_\bk$ account for particle-hole (p-h)
asymmetry present in  Cd$_3$As$_2$.

Following Refs. \cite{shapourian2015,cortijo2015,fujimoto2016,pikulin2016} the most
important  effect of elastic strain can be included in the lattice
model (\ref{h1}) by modifying the
electron tunneling amplitude along the $\hat{z}$-direction according to
\bee \label{h4}   
t_1\tau^z\to t_1(1-u_{33})\tau^z+i\Lambda\sum_{j\neq 3}u_{3j}\tau^j,
\ee
where $u_{ij}={1\over 2}(\partial_iu_j+\partial_ju_i)$ is the
symmetrized strain tensor and $\bu=(u_1,u_2,u_3)$ represents the
displacement of the atoms. To see how this leads to an emergent vector
potential we study the low-energy effective
theory. We expand $h^{\rm latt}(\bk)$  in the vicinity of the
Weyl points $\bK_\pm$ by writing $\bk=\bK_\pm +\bq$  and assuming
small $|\bq|$. To leading order we obtain the linearized Hamiltonian
of the distorted crystal \cite{pikulin2016}
\bee \label{h6}
h_{\eta}(\bq)= v_{\eta} ^j\tau^j\left(\hbar q_j-\eta{e\over c}\cA_j\right),
\ee
with the velocity vector 
\bee \label{h3cc}
\bv_{\eta}=\hbar^{-1}a(\Lambda, \Lambda, -\eta t_1 \sin{aQ}).
\ee
 For Cd$_2$As$_3$ parameters and lattice constant $a=4$\AA \ this gives $\hbar\bv_{\eta}=(0.89,0.89,-1.24\eta)$eV\AA.
The strain-induced gauge potential is given by
\bee \label{h7}
\vec\cA=-{\hbar c\over ea}\bigl(u_{13}\sin{aQ},u_{23}\sin{aQ},u_{33}\cot{aQ}\bigr).
\ee
\begin{figure*}[t]
\includegraphics[width = 15.5cm]{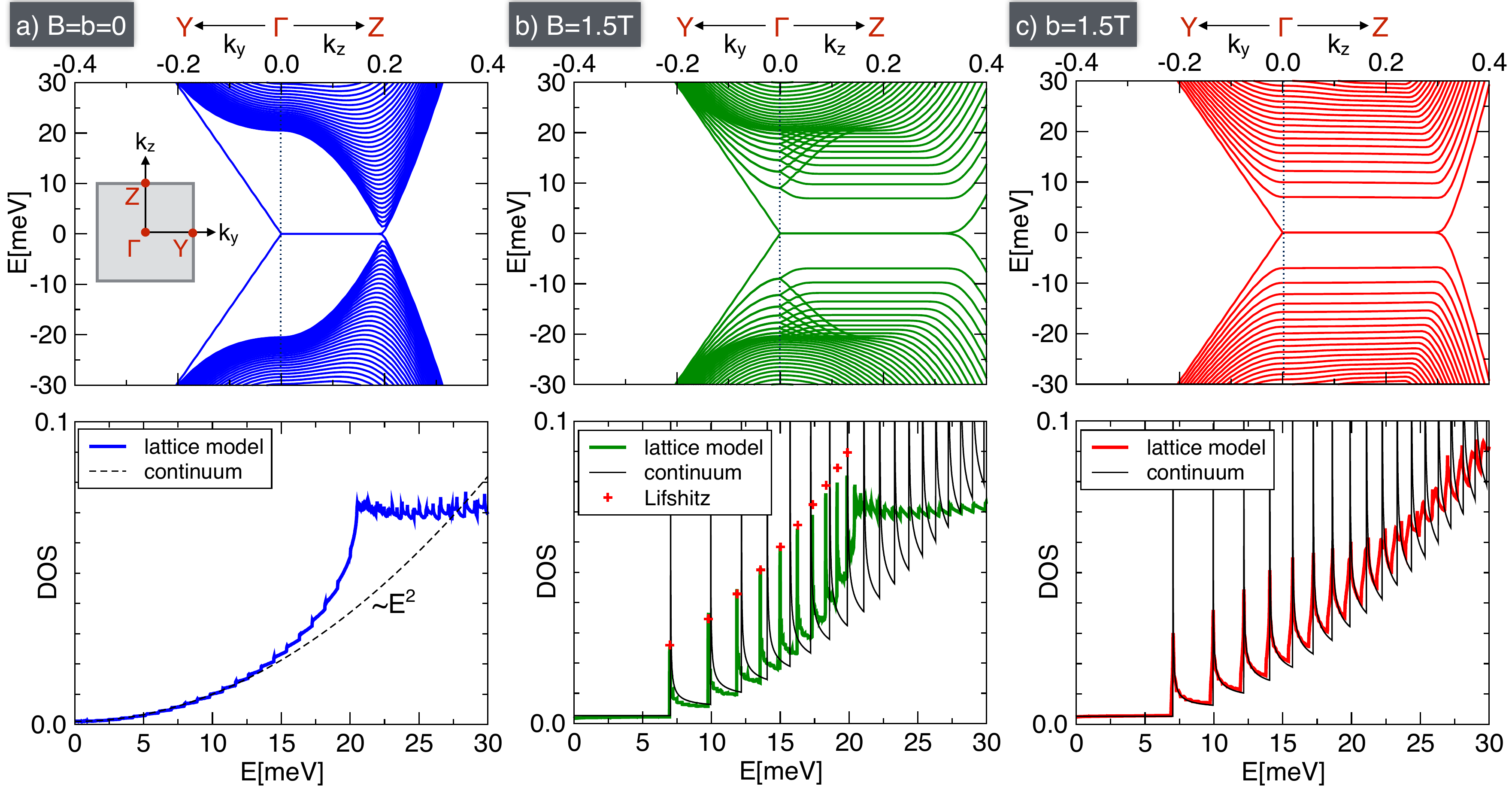}
\caption{Numerical results for the Cd$_3$As$_2$ lattice Hamiltonian
  (\ref{h3}) in the presence of magnetic field $\bB=\hat{y}B$ and
  strain-induced pseudomagnetic field $\bb=\hat{y}b$. In all
  panels films of thickness 500 lattice points are studied with
  parameters appropriate for Cd$_3$As$_2$. P-h asymmetry terms
  $\epsilon_\bk$ are neglected for simplicity which makes
  contributions from the two spin sectors identical. a) Band structure and
  density of states (DOS) for zero field and zero strain. The inset shows
  the first Brillouin zone. b)  Band structure and normalized DOS
  for $B=1.5$T. Red crosses indicate the peak positions expected on the
  basis of the Lifshitz-Onsager quantization condition \cite{shoenberg_book}. c) Band structure and DOS
  for $b=1.5$T. Thin black line shows the expected bulk DOS for ideal
  Weyl dispersion computed from Eq.\ (\ref{lan1}). 
}\label{fig2}
\end{figure*}
We see that elements $u_{j3}$ of the strain tensor act on the
low-energy Weyl fermions as components of a chiral gauge field
 because according to Eq. (\ref{h6}) $\vec\cA$ couples with
the opposite sign to the Weyl fermions with opposite chirality
$\eta$. Ordinary electromagnetic gauge potential couples through the 
replacement $\hbar\bq\to \hbar\bq-{e\over c}\bA$, independent of
$\eta$. Ref.\ \cite{pikulin2016} noted that application of a torsional
strain to a nanowire made of Cd$_3$As$_2$ (grown along the 001
crystallographic direction) results in a uniform pseudomagnetic field
$\bb=\nabla\times\vec\cA$ pointed along the axis of the wire. The strength
of this pseudomagnetic field was estimated as $b\lesssim 0.3$T which would
be insufficient to observe QO. Our key observation
here is that a different type of distortion, illustrated in Fig.\
\ref{fig1}a,  can produce a much larger field $b$.

One reason why the torsion-induced $b$-field is relatively small lies
in the fact that it originates from the $\cA_x$ and $\cA_y$ components of
the vector potential. According to Eq.\ (\ref{h7}) these are
suppressed relative to the strain components by a factor of $\sin{aQ}$.
This is a small number in most Dirac and Weyl semimetals because the
distance  $2Q$ between the Weyl points is typically a small fraction
of the Brillouin zone size $2\pi/a$. Specifically, we have $aQ\simeq 0.132$
in Cd$_3$As$_2$  \cite{zhizhun2013}.  Note on the other hand that the $\cA_z$ component of
the chiral gauge potential comes with a factor $\cot{aQ}\simeq 1/aQ$
and is therefore enhanced. A lattice distortion that
produces nonzero strain tensor element $u_{33}$ will therefore be much more efficient 
in generating large $b$ than $u_{13}$ or $u_{23}$. Specifically, for
the same amount of strain the field strength is enhanced by a factor
of $\cot{aQ}/\sin{aQ}\simeq 1/(aQ)^2\simeq 57$ for  Cd$_3$As$_2$.

To implement this type of strain we consider a thin film (or a
nanowire) grown such that vector $\bK_\eta$ lies along the $z$ direction
as defined in Fig.\ \ref{fig1}a. More generally we require that
$\bK_\eta$  has a nonzero projection onto
the surface of the film or on the long direction for the nanowire.
Cd$_3$As$_2$ films \cite{liu2015}, microribbons \cite{hui2016}
and nanowires \cite{li2015,wang2016} satisfy this requirement. Bending
the film as shown in Fig.\ \ref{fig1}b creates a displacement field
$\bu=(0,0,2\alpha xz/d)$, where $d$ is the film thickness and $\alpha$
controls the magnitude of the bend. (If $R$ is the radius of the
circular section formed by the bent film then $\alpha=2d/R$. $\alpha$
can also be interpreted as the maximum fractional displacement
$\alpha=u_{\rm max}/a$ that occurs at the surface of the film.) This
distortion gives $u_{33}=2\alpha x/d$
which, through Eq.\ (\ref{h7}), yields a pseudomagnetic field 
\bee \label{h8}
\bb=\nabla\times\vec\cA=\hat{y}\left({2\alpha\over d}\right){\hbar c\over ea}\cot{aQ}.
\ee
Noting that $\Phi_0=hc/e=4.12\times10^5$T\AA \ we may estimate the
resulting field strength for a $d=100$nm film as 
\bee \label{h9}
b\simeq\alpha\times 246{\rm T}.
\ee
The maximum pseudomagnetic field that can be achieved will depend on
the maximum strain that the material can sustain. Ref.\ \cite{li2015}
characterized the Cd$_3$As$_2$ nanowires as ``greatly flexible'' and their Figure
1a shows some wires bent with a radius $R$ as small as several microns. This implies
that $\alpha$ of several percent can likely be achieved. From Eq.\ (\ref{h9}) we thus estimate that field
strength $b\simeq 10-15$T can be reached, providing a substantial
window for the observation of the strain-induced QO.

To substantiate these claims we now present the results of our
numerical simulations based on the lattice Hamiltonian
(\ref{h3}). Magnetic field $B$ is
implemented via the standard Peierls substitution while the
strain-induced field $b$ through
Eq.\ (\ref{h4}). Geometry outlined in Fig.\ \ref{fig1} is used with periodic boundary
conditions along $y$ and $z$, open along $x$.  Fig.\ \ref{fig2} provides the summary of our results.
The unstrained crystal at zero field (panel a) shows the expected band
structure with bulk Weyl nodes close to $k_za=\pm 0.2$ and a pair of
linearly dispersing surface states corresponding to Fermi arcs. The
density of states (DOS) exhibits the expected quadratic behavior
$D(E)\sim E^2$ at low energies with some deviations
apparent for $|E| \gtrsim 12$meV due to the departure of the lattice
model from the
perfectly linear Weyl dispersion. At $E_{\rm Lif}\simeq 20$meV Lifshitz
transition occurs where two small Fermi surfaces associated with each
Weyl point merge into a single large Fermi surface as illustrated in
Fig.\ \ref{fig0}b.  

In Fig.\ \ref{fig2}b magnetic field $\bB=\hat{y}B$ is seen to reorganize the linearly
dispersing bulk bands into flat Landau levels. In the
continuum approximation given by Eq.\ (\ref{h6}) the bulk spectrum of such
Dirac-Landau levels is well known and reads
\bee \label{lan1}
E_{n}(k_y)=\pm \hbar \sqrt{v_y^2k_y^2+2nv_xv_z{e|B|\over \hbar c}}, \ \ \ n=1,2,\dots,
\ee
The corresponding DOS shows a series of spikes at the onset of each
new Landau level and is in a good agreement with the DOS
calculated from the lattice model. Deviations occur above $\sim 12$meV
because the energy dispersion of the lattice model is no longer perfectly linear at higher
energies. The peak positions $E_n$ agree perfectly with the Lifshitz-Onsager
quantization condition \cite{shoenberg_book},  which takes into account
these deviations. It requires that   $S(E_n)=2\pi
  n(eB/\hbar c)$, where $S(E)$ is the extremal cross-sectional
  area of a surface of constant energy $E$ in the plane perpendicular
  to $\bB$ (see Fig.\ \ref{fig0}b), and $n=1,2,\cdots$.

Pseudomagnetic field $\bb=\hat{y}b$, induced by strain using Eq.\
(\ref{h4}) with $u_{33}=2\alpha x/d$, also generates flat bands
(panel c), as expected on the basis of arguments presented above. The
corresponding DOS is in agreement with that obtained from Eq.\
(\ref{lan1}) upon replacing $B\to b$. Remarkably the agreement is
nearly perfect for all energies up to $E_{\rm Lif}$. We attribute this interesting result to the fact that strain
couples as the chiral vector potential  {\em only} to the
Weyl fermions. If we write the full Hamiltonian as
$h(\bp)=h_W(\bp) +\delta h(\bp)$ where $h_W$ is strictly linear in momentum
$\bp$ and $\delta h$ is the correction resulting from the lattice
effects, then strain causes $\bp\to \bp-{e\over c}\vec\cA$ only in $h_W$
but does not to leading order affect $\delta h$. The real vector
potential $\bA$ affects  $h_W$ and  $\delta h$ in the same way.

\begin{figure}[t]
\includegraphics[width = 8.0cm]{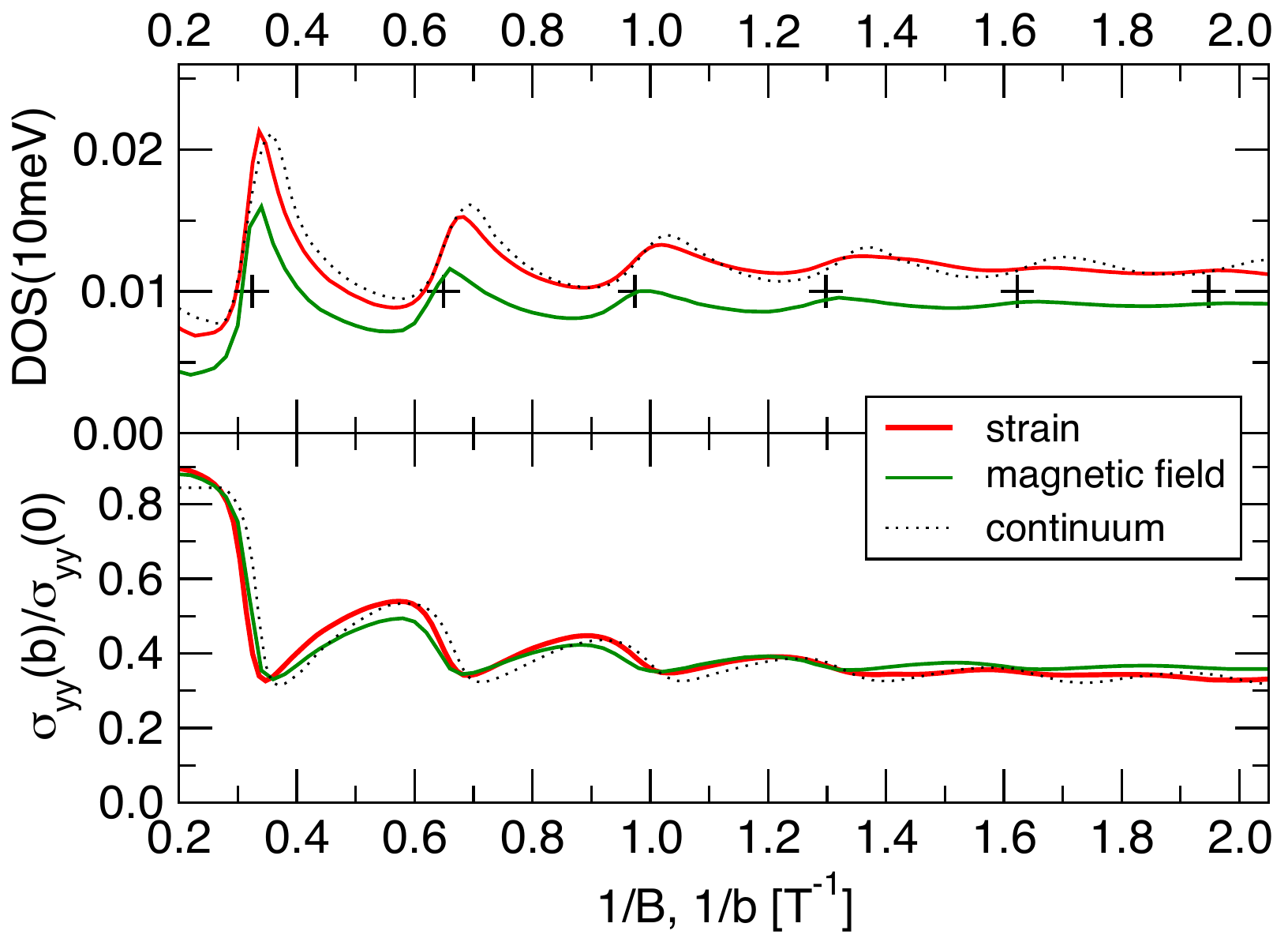}
\caption{Strain-induced QO. Top pannel shows
  oscillations in DOS at energy 10meV as a function of inverse strain
  strength expressed as $1/b$. For comparison ordinary magnetic
  oscillations are displayed, as well as the result of the
  bulk continuum theory Eq.\ (\ref{lan1}). Crosses indicate peak positions expected based on the  Lifshitz-Onsager theory.  Bottom pannel shows
  oscillations in conductivity $\sigma_{yy}$ assuming Fermi energy
  $E_F=10$meV. To simulate the effect of disorder all data are
  broadened by convolving in energy with a Lorentzian with width $\delta=0.25$meV. 
The same geometry and parameters are used
  as in Fig.\ \ref{fig2}.
}\label{fig3}
\end{figure}

These results imply that QO will occur when either
$B$ or $b$ is present. If we  vary
$B$ then $D(E_F)$, together with most measurable quantities, will
exhibit oscillations periodic in $1/B$. The same is true for the
strain-induced pseudomagnetic field $b$. This is illustrated in Fig.\
\ref{fig3} which shows oscillations in DOS and longitudinal
conductivity $\sigma_{yy}$ at energy 10meV as a function of $1/b$ and
$1/B$.  Conductivity is calculated using the standard relaxation time
approximation as described in SM.
Strain-induced QO show robust periodicity in
$1/b$. Their period 0.329T$^{-1}$ is in a good agreement with the
period 0.324T$^{-1}$ expected on the basis of the Lifshitz-Onsager
theory and 0.336T$^{-1}$ obtained from Eq.\ (\ref{lan1}). Small
irregularities that appear at low fields can be attributed
 to the finite size effects as the Landau level spacing  becomes
 comparable to the subband spacing apparent e.g.\ in Fig.\ \ref{fig2}a.
We verified that similar oscillations occur at other energies below
the Lifshitz transition. Remarkably, we find strain-induced
oscillations periodic in $1/b$ also above $E_{\rm Lif}$.  In addition, we expect that in the presence of both $b$ and $B$ fields the peaks
split as two Weyl cones feel different effective magnetic
fields. These effects are further discussed in SM.

Results presented above extend trivially to the full Cd$_3$As$_2$ Hamiltonian
Eq.\ (\ref{h1}) where the spin-down block makes an identical
contribution and the p-h symmetry breaking terms contained
in $\epsilon_\bk$ bring only quantitative changes (see SM for
discussion).  Experimental studies
\cite{borisenko2014,neupane2014,jeon2014,liu2014b} indicate that the
linear dispersion in Cd$_3$As$_2$ extends over a much wider range
of energies than theoretically anticipated \cite{zhizhun2013}
with the Lifshitz transition occurring above $\sim 200$meV. We therefore expect the
zero-field strain-induced QO predicted in this work 
to be easily observable in suitably fabricated Cd$_3$As$_2$ films and
nanowires and potentially also in other Dirac and Weyl semimetals. Our
results show that conditions for their observability are identical to
those required to detect ordinary QO. The
continuous tunability of the pseudomagnetic field in large parameter range provides a new experimental basis for the
study of emergent gauge fields in three-dimensional crystalline solids.

\begin{acknowledgments}

The authors are indebted to D.A. Bonn, D.M. Broun, A. Chen, I. Elfimov
and W. N. Hardy for illuminating discussions, and thank NSERC, CIfAR and Max Planck - UBC Centre for Quantum Materials for support.
\end{acknowledgments}

%...
\bibliography{weyl2.bib}

\begin{thebibliography}{47}
\expandafter\ifx\csname natexlab\endcsname\relax\def\natexlab#1{#1}\fi
\expandafter\ifx\csname bibnamefont\endcsname\relax
  \def\bibnamefont#1{#1}\fi
\expandafter\ifx\csname bibfnamefont\endcsname\relax
  \def\bibfnamefont#1{#1}\fi
\expandafter\ifx\csname citenamefont\endcsname\relax
  \def\citenamefont#1{#1}\fi
\expandafter\ifx\csname url\endcsname\relax
  \def\url#1{\texttt{#1}}\fi
\expandafter\ifx\csname urlprefix\endcsname\relax\def\urlprefix{URL }\fi
\providecommand{\bibinfo}[2]{#2}
\providecommand{\eprint}[2][]{\url{#2}}

\bibitem[{\citenamefont{Wan et~al.}(2011)\citenamefont{Wan, Turner, Vishwanath,
  and Savrasov}}]{Savrasov2011}
\bibinfo{author}{\bibfnamefont{X.}~\bibnamefont{Wan}},
  \bibinfo{author}{\bibfnamefont{A.~M.} \bibnamefont{Turner}},
  \bibinfo{author}{\bibfnamefont{A.}~\bibnamefont{Vishwanath}},
  \bibnamefont{and} \bibinfo{author}{\bibfnamefont{S.~Y.}
  \bibnamefont{Savrasov}}, \bibinfo{journal}{Phys. Rev. B}
  \textbf{\bibinfo{volume}{83}}, \bibinfo{pages}{205101}
  (\bibinfo{year}{2011}).

\bibitem[{\citenamefont{Burkov et~al.}(2011)\citenamefont{Burkov, Hook, and
  Balents}}]{burkov2011b}
\bibinfo{author}{\bibfnamefont{A.~A.} \bibnamefont{Burkov}},
  \bibinfo{author}{\bibfnamefont{M.~D.} \bibnamefont{Hook}}, \bibnamefont{and}
  \bibinfo{author}{\bibfnamefont{L.}~\bibnamefont{Balents}},
  \bibinfo{journal}{Phys. Rev. B} \textbf{\bibinfo{volume}{84}},
  \bibinfo{pages}{235126} (\bibinfo{year}{2011}).

\bibitem[{\citenamefont{Vafek and Vishwanath}(2014)}]{Vafek2014}
\bibinfo{author}{\bibfnamefont{O.}~\bibnamefont{Vafek}} \bibnamefont{and}
  \bibinfo{author}{\bibfnamefont{A.}~\bibnamefont{Vishwanath}},
  \bibinfo{journal}{Annual Review of Condensed Matter Physics}
  \textbf{\bibinfo{volume}{5}}, \bibinfo{pages}{83} (\bibinfo{year}{2014}).

\bibitem[{\citenamefont{Fukushima et~al.}(2008)\citenamefont{Fukushima,
  Kharzeev, and Warringa}}]{fukushima2008}
\bibinfo{author}{\bibfnamefont{K.}~\bibnamefont{Fukushima}},
  \bibinfo{author}{\bibfnamefont{D.~E.} \bibnamefont{Kharzeev}},
  \bibnamefont{and} \bibinfo{author}{\bibfnamefont{H.~J.}
  \bibnamefont{Warringa}}, \bibinfo{journal}{Phys. Rev. D}
  \textbf{\bibinfo{volume}{78}}, \bibinfo{pages}{074033}
  (\bibinfo{year}{2008}).

\bibitem[{\citenamefont{Son and Spivak}(2013)}]{son2013}
\bibinfo{author}{\bibfnamefont{D.~T.} \bibnamefont{Son}} \bibnamefont{and}
  \bibinfo{author}{\bibfnamefont{B.~Z.} \bibnamefont{Spivak}},
  \bibinfo{journal}{Phys. Rev. B} \textbf{\bibinfo{volume}{88}},
  \bibinfo{pages}{104412} (\bibinfo{year}{2013}).

\bibitem[{\citenamefont{Kim et~al.}(2013)\citenamefont{Kim, Kim, Wang, Sasaki,
  Satoh, Ohnishi, Kitaura, Yang, and Li}}]{kim2013}
\bibinfo{author}{\bibfnamefont{H.-J.} \bibnamefont{Kim}},
  \bibinfo{author}{\bibfnamefont{K.-S.} \bibnamefont{Kim}},
  \bibinfo{author}{\bibfnamefont{J.-F.} \bibnamefont{Wang}},
  \bibinfo{author}{\bibfnamefont{M.}~\bibnamefont{Sasaki}},
  \bibinfo{author}{\bibfnamefont{N.}~\bibnamefont{Satoh}},
  \bibinfo{author}{\bibfnamefont{A.}~\bibnamefont{Ohnishi}},
  \bibinfo{author}{\bibfnamefont{M.}~\bibnamefont{Kitaura}},
  \bibinfo{author}{\bibfnamefont{M.}~\bibnamefont{Yang}}, \bibnamefont{and}
  \bibinfo{author}{\bibfnamefont{L.}~\bibnamefont{Li}}, \bibinfo{journal}{Phys.
  Rev. Lett.} \textbf{\bibinfo{volume}{111}}, \bibinfo{pages}{246603}
  (\bibinfo{year}{2013}).

\bibitem[{\citenamefont{Huang et~al.}(2015)\citenamefont{Huang, Zhao, Long,
  Wang, Chen, Yang, Liang, Xue, Weng, Fang et~al.}}]{huang2015}
\bibinfo{author}{\bibfnamefont{X.}~\bibnamefont{Huang}},
  \bibinfo{author}{\bibfnamefont{L.}~\bibnamefont{Zhao}},
  \bibinfo{author}{\bibfnamefont{Y.}~\bibnamefont{Long}},
  \bibinfo{author}{\bibfnamefont{P.}~\bibnamefont{Wang}},
  \bibinfo{author}{\bibfnamefont{D.}~\bibnamefont{Chen}},
  \bibinfo{author}{\bibfnamefont{Z.}~\bibnamefont{Yang}},
  \bibinfo{author}{\bibfnamefont{H.}~\bibnamefont{Liang}},
  \bibinfo{author}{\bibfnamefont{M.}~\bibnamefont{Xue}},
  \bibinfo{author}{\bibfnamefont{H.}~\bibnamefont{Weng}},
  \bibinfo{author}{\bibfnamefont{Z.}~\bibnamefont{Fang}}, \bibnamefont{et~al.},
  \bibinfo{journal}{Phys. Rev. X} \textbf{\bibinfo{volume}{5}},
  \bibinfo{pages}{031023} (\bibinfo{year}{2015}).

\bibitem[{\citenamefont{Xiong et~al.}(2015)\citenamefont{Xiong, Kushwaha,
  Liang, Krizan, Hirschberger, Wang, Cava, and Ong}}]{ong2015}
\bibinfo{author}{\bibfnamefont{J.}~\bibnamefont{Xiong}},
  \bibinfo{author}{\bibfnamefont{S.~K.} \bibnamefont{Kushwaha}},
  \bibinfo{author}{\bibfnamefont{T.}~\bibnamefont{Liang}},
  \bibinfo{author}{\bibfnamefont{J.~W.} \bibnamefont{Krizan}},
  \bibinfo{author}{\bibfnamefont{M.}~\bibnamefont{Hirschberger}},
  \bibinfo{author}{\bibfnamefont{W.}~\bibnamefont{Wang}},
  \bibinfo{author}{\bibfnamefont{R.~J.} \bibnamefont{Cava}}, \bibnamefont{and}
  \bibinfo{author}{\bibfnamefont{N.~P.} \bibnamefont{Ong}},
  \bibinfo{journal}{Science} \textbf{\bibinfo{volume}{350}},
  \bibinfo{pages}{413} (\bibinfo{year}{2015}).

\bibitem[{\citenamefont{Burkov}(2015)}]{burkov2015}
\bibinfo{author}{\bibfnamefont{A.~A.} \bibnamefont{Burkov}},
  \bibinfo{journal}{Journal of Physics: Condensed Matter}
  \textbf{\bibinfo{volume}{27}}, \bibinfo{pages}{113201}
  (\bibinfo{year}{2015}).

\bibitem[{\citenamefont{Li et~al.}(2016{\natexlab{a}})\citenamefont{Li,
  Kharzeev, Zhang, Huang, Pletikosic, Fedorov, Zhong, Schneeloch, Gu, and
  Valla}}]{valla2016}
\bibinfo{author}{\bibfnamefont{Q.}~\bibnamefont{Li}},
  \bibinfo{author}{\bibfnamefont{D.~E.} \bibnamefont{Kharzeev}},
  \bibinfo{author}{\bibfnamefont{C.}~\bibnamefont{Zhang}},
  \bibinfo{author}{\bibfnamefont{Y.}~\bibnamefont{Huang}},
  \bibinfo{author}{\bibfnamefont{I.}~\bibnamefont{Pletikosic}},
  \bibinfo{author}{\bibfnamefont{A.~V.} \bibnamefont{Fedorov}},
  \bibinfo{author}{\bibfnamefont{R.~D.} \bibnamefont{Zhong}},
  \bibinfo{author}{\bibfnamefont{J.~A.} \bibnamefont{Schneeloch}},
  \bibinfo{author}{\bibfnamefont{G.~D.} \bibnamefont{Gu}}, \bibnamefont{and}
  \bibinfo{author}{\bibfnamefont{T.}~\bibnamefont{Valla}},
  \bibinfo{journal}{Nat Phys} \textbf{\bibinfo{volume}{12}},
  \bibinfo{pages}{550} (\bibinfo{year}{2016}{\natexlab{a}}).

\bibitem[{\citenamefont{Zhang et~al.}(2016)\citenamefont{Zhang, Xu, Belopolski,
  Yuan, Lin, Tong, Bian, Alidoust, Lee, Huang et~al.}}]{jia2016}
\bibinfo{author}{\bibfnamefont{C.-L.} \bibnamefont{Zhang}},
  \bibinfo{author}{\bibfnamefont{S.-Y.} \bibnamefont{Xu}},
  \bibinfo{author}{\bibfnamefont{I.}~\bibnamefont{Belopolski}},
  \bibinfo{author}{\bibfnamefont{Z.}~\bibnamefont{Yuan}},
  \bibinfo{author}{\bibfnamefont{Z.}~\bibnamefont{Lin}},
  \bibinfo{author}{\bibfnamefont{B.}~\bibnamefont{Tong}},
  \bibinfo{author}{\bibfnamefont{G.}~\bibnamefont{Bian}},
  \bibinfo{author}{\bibfnamefont{N.}~\bibnamefont{Alidoust}},
  \bibinfo{author}{\bibfnamefont{C.-C.} \bibnamefont{Lee}},
  \bibinfo{author}{\bibfnamefont{S.-M.} \bibnamefont{Huang}},
  \bibnamefont{et~al.}, \bibinfo{journal}{Nat Commun}
  \textbf{\bibinfo{volume}{7}}, \bibinfo{pages}{10735} (\bibinfo{year}{2016}).

\bibitem[{\citenamefont{Adler}(1969)}]{adler1969}
\bibinfo{author}{\bibfnamefont{S.~L.} \bibnamefont{Adler}},
  \bibinfo{journal}{Phys. Rev.} \textbf{\bibinfo{volume}{177}},
  \bibinfo{pages}{2426} (\bibinfo{year}{1969}).

\bibitem[{\citenamefont{Bell and Jackiw}(1969)}]{bell1969}
\bibinfo{author}{\bibfnamefont{J.~S.} \bibnamefont{Bell}} \bibnamefont{and}
  \bibinfo{author}{\bibfnamefont{R.}~\bibnamefont{Jackiw}},
  \bibinfo{journal}{Il Nuovo Cimento A (1971-1996)}
  \textbf{\bibinfo{volume}{60}}, \bibinfo{pages}{47} (\bibinfo{year}{1969}).

\bibitem[{\citenamefont{Nielsen and Ninomiya}(1983)}]{nielsen1983}
\bibinfo{author}{\bibfnamefont{H.}~\bibnamefont{Nielsen}} \bibnamefont{and}
  \bibinfo{author}{\bibfnamefont{M.}~\bibnamefont{Ninomiya}},
  \bibinfo{journal}{Physics Letters B} \textbf{\bibinfo{volume}{130}},
  \bibinfo{pages}{389 } (\bibinfo{year}{1983}).

\bibitem[{\citenamefont{Parameswaran et~al.}(2014)\citenamefont{Parameswaran,
  Grover, Abanin, Pesin, and Vishwanath}}]{parameswaran2014}
\bibinfo{author}{\bibfnamefont{S.~A.} \bibnamefont{Parameswaran}},
  \bibinfo{author}{\bibfnamefont{T.}~\bibnamefont{Grover}},
  \bibinfo{author}{\bibfnamefont{D.~A.} \bibnamefont{Abanin}},
  \bibinfo{author}{\bibfnamefont{D.~A.} \bibnamefont{Pesin}}, \bibnamefont{and}
  \bibinfo{author}{\bibfnamefont{A.}~\bibnamefont{Vishwanath}},
  \bibinfo{journal}{Phys. Rev. X} \textbf{\bibinfo{volume}{4}},
  \bibinfo{pages}{031035} (\bibinfo{year}{2014}).

\bibitem[{\citenamefont{Baum et~al.}(2015)\citenamefont{Baum, Berg,
  Parameswaran, and Stern}}]{baum2016}
\bibinfo{author}{\bibfnamefont{Y.}~\bibnamefont{Baum}},
  \bibinfo{author}{\bibfnamefont{E.}~\bibnamefont{Berg}},
  \bibinfo{author}{\bibfnamefont{S.~A.} \bibnamefont{Parameswaran}},
  \bibnamefont{and} \bibinfo{author}{\bibfnamefont{A.}~\bibnamefont{Stern}},
  \bibinfo{journal}{Phys. Rev. X} \textbf{\bibinfo{volume}{5}},
  \bibinfo{pages}{041046} (\bibinfo{year}{2015}).

\bibitem[{\citenamefont{Chen and Franz}(2016)}]{anffany2016}
\bibinfo{author}{\bibfnamefont{A.}~\bibnamefont{Chen}} \bibnamefont{and}
  \bibinfo{author}{\bibfnamefont{M.}~\bibnamefont{Franz}},
  \bibinfo{journal}{Phys. Rev. B} \textbf{\bibinfo{volume}{93}},
  \bibinfo{pages}{201105} (\bibinfo{year}{2016}).

\bibitem[{\citenamefont{Potter et~al.}(2014)\citenamefont{Potter, Kimchi, and
  Vishwanath}}]{Potter2014}
\bibinfo{author}{\bibfnamefont{A.~C.} \bibnamefont{Potter}},
  \bibinfo{author}{\bibfnamefont{I.}~\bibnamefont{Kimchi}}, \bibnamefont{and}
  \bibinfo{author}{\bibfnamefont{A.}~\bibnamefont{Vishwanath}},
  \bibinfo{journal}{Nat Commun} \textbf{\bibinfo{volume}{5}},
  \bibinfo{pages}{5161} (\bibinfo{year}{2014}).

\bibitem[{\citenamefont{Moll et~al.}(2016)\citenamefont{Moll, Nair, Helm,
  Potter, Kimchi, Vishwanath, and Analytis}}]{Moll2016}
\bibinfo{author}{\bibfnamefont{P.~J.~W.} \bibnamefont{Moll}},
  \bibinfo{author}{\bibfnamefont{N.~L.} \bibnamefont{Nair}},
  \bibinfo{author}{\bibfnamefont{T.}~\bibnamefont{Helm}},
  \bibinfo{author}{\bibfnamefont{A.~C.} \bibnamefont{Potter}},
  \bibinfo{author}{\bibfnamefont{I.}~\bibnamefont{Kimchi}},
  \bibinfo{author}{\bibfnamefont{A.}~\bibnamefont{Vishwanath}},
  \bibnamefont{and} \bibinfo{author}{\bibfnamefont{J.~G.}
  \bibnamefont{Analytis}}, \bibinfo{journal}{Nature}
  \textbf{\bibinfo{volume}{535}}, \bibinfo{pages}{266} (\bibinfo{year}{2016}).

\bibitem[{\citenamefont{Shoenberg}(1984)}]{shoenberg_book}
\bibinfo{author}{\bibfnamefont{D.}~\bibnamefont{Shoenberg}},
  \emph{\bibinfo{title}{Magnetic Oscillations in Metals}}
  (\bibinfo{publisher}{Cambridge University Press, Cambridge},
  \bibinfo{year}{1984}).

\bibitem[{\citenamefont{Guinea et~al.}(2010)\citenamefont{Guinea, Katsnelson,
  and Geim}}]{guinea2010}
\bibinfo{author}{\bibfnamefont{F.}~\bibnamefont{Guinea}},
  \bibinfo{author}{\bibfnamefont{M.~I.} \bibnamefont{Katsnelson}},
  \bibnamefont{and} \bibinfo{author}{\bibfnamefont{A.~K.} \bibnamefont{Geim}},
  \bibinfo{journal}{Nat Phys} \textbf{\bibinfo{volume}{6}}, \bibinfo{pages}{30}
  (\bibinfo{year}{2010}).

\bibitem[{\citenamefont{Levy et~al.}(2010)\citenamefont{Levy, Burke, Meaker,
  Panlasigui, Zettl, Guinea, Neto, and Crommie}}]{levy2010}
\bibinfo{author}{\bibfnamefont{N.}~\bibnamefont{Levy}},
  \bibinfo{author}{\bibfnamefont{S.~A.} \bibnamefont{Burke}},
  \bibinfo{author}{\bibfnamefont{K.~L.} \bibnamefont{Meaker}},
  \bibinfo{author}{\bibfnamefont{M.}~\bibnamefont{Panlasigui}},
  \bibinfo{author}{\bibfnamefont{A.}~\bibnamefont{Zettl}},
  \bibinfo{author}{\bibfnamefont{F.}~\bibnamefont{Guinea}},
  \bibinfo{author}{\bibfnamefont{A.~H.~C.} \bibnamefont{Neto}},
  \bibnamefont{and} \bibinfo{author}{\bibfnamefont{M.~F.}
  \bibnamefont{Crommie}}, \bibinfo{journal}{Science}
  \textbf{\bibinfo{volume}{329}}, \bibinfo{pages}{544} (\bibinfo{year}{2010}).

\bibitem[{\citenamefont{Shapourian et~al.}(2015)\citenamefont{Shapourian,
  Hughes, and Ryu}}]{shapourian2015}
\bibinfo{author}{\bibfnamefont{H.}~\bibnamefont{Shapourian}},
  \bibinfo{author}{\bibfnamefont{T.~L.} \bibnamefont{Hughes}},
  \bibnamefont{and} \bibinfo{author}{\bibfnamefont{S.}~\bibnamefont{Ryu}},
  \bibinfo{journal}{Phys. Rev. B} \textbf{\bibinfo{volume}{92}},
  \bibinfo{pages}{165131} (\bibinfo{year}{2015}).

\bibitem[{\citenamefont{Cortijo et~al.}(2015)\citenamefont{Cortijo,
  Ferreir\'os, Landsteiner, and Vozmediano}}]{cortijo2015}
\bibinfo{author}{\bibfnamefont{A.}~\bibnamefont{Cortijo}},
  \bibinfo{author}{\bibfnamefont{Y.}~\bibnamefont{Ferreir\'os}},
  \bibinfo{author}{\bibfnamefont{K.}~\bibnamefont{Landsteiner}},
  \bibnamefont{and} \bibinfo{author}{\bibfnamefont{M.~A.~H.}
  \bibnamefont{Vozmediano}}, \bibinfo{journal}{Phys. Rev. Lett.}
  \textbf{\bibinfo{volume}{115}}, \bibinfo{pages}{177202}
  (\bibinfo{year}{2015}).

\bibitem[{\citenamefont{Sumiyoshi and Fujimoto}(2016)}]{fujimoto2016}
\bibinfo{author}{\bibfnamefont{H.}~\bibnamefont{Sumiyoshi}} \bibnamefont{and}
  \bibinfo{author}{\bibfnamefont{S.}~\bibnamefont{Fujimoto}},
  \bibinfo{journal}{Phys. Rev. Lett.} \textbf{\bibinfo{volume}{116}},
  \bibinfo{pages}{166601} (\bibinfo{year}{2016}).

\bibitem[{\citenamefont{Pikulin et~al.}(unpublished)\citenamefont{Pikulin,
  Chen, and Franz}}]{pikulin2016}
\bibinfo{author}{\bibfnamefont{D.}~\bibnamefont{Pikulin}},
  \bibinfo{author}{\bibfnamefont{A.}~\bibnamefont{Chen}}, \bibnamefont{and}
  \bibinfo{author}{\bibfnamefont{M.}~\bibnamefont{Franz}},
  \bibinfo{journal}{arXiv:1607.01810}  (\bibinfo{year}{unpublished}).

\bibitem[{\citenamefont{{Grushin} et~al.}(2016)\citenamefont{{Grushin},
  {Venderbos}, {Vishwanath}, and {Ilan}}}]{Grushin2016}
\bibinfo{author}{\bibfnamefont{A.~G.} \bibnamefont{{Grushin}}},
  \bibinfo{author}{\bibfnamefont{J.~W.~F.} \bibnamefont{{Venderbos}}},
  \bibinfo{author}{\bibfnamefont{A.}~\bibnamefont{{Vishwanath}}},
  \bibnamefont{and} \bibinfo{author}{\bibfnamefont{R.}~\bibnamefont{{Ilan}}},
  \bibinfo{journal}{ArXiv e-prints}  (\bibinfo{year}{2016}),
  \eprint{1607.04268}.

\bibitem[{\citenamefont{He et~al.}(2014)\citenamefont{He, Hong, Dong, Pan,
  Zhang, Zhang, and Li}}]{he2014}
\bibinfo{author}{\bibfnamefont{L.~P.} \bibnamefont{He}},
  \bibinfo{author}{\bibfnamefont{X.~C.} \bibnamefont{Hong}},
  \bibinfo{author}{\bibfnamefont{J.~K.} \bibnamefont{Dong}},
  \bibinfo{author}{\bibfnamefont{J.}~\bibnamefont{Pan}},
  \bibinfo{author}{\bibfnamefont{Z.}~\bibnamefont{Zhang}},
  \bibinfo{author}{\bibfnamefont{J.}~\bibnamefont{Zhang}}, \bibnamefont{and}
  \bibinfo{author}{\bibfnamefont{S.~Y.} \bibnamefont{Li}},
  \bibinfo{journal}{Phys. Rev. Lett.} \textbf{\bibinfo{volume}{113}},
  \bibinfo{pages}{246402} (\bibinfo{year}{2014}).

\bibitem[{\citenamefont{Liu et~al.}(2015)\citenamefont{Liu, Zhang, Yuan, Lei,
  Wang, Di~Sante, Narayan, He, Picozzi, Sanvito et~al.}}]{liu2015}
\bibinfo{author}{\bibfnamefont{Y.}~\bibnamefont{Liu}},
  \bibinfo{author}{\bibfnamefont{C.}~\bibnamefont{Zhang}},
  \bibinfo{author}{\bibfnamefont{X.}~\bibnamefont{Yuan}},
  \bibinfo{author}{\bibfnamefont{T.}~\bibnamefont{Lei}},
  \bibinfo{author}{\bibfnamefont{C.}~\bibnamefont{Wang}},
  \bibinfo{author}{\bibfnamefont{D.}~\bibnamefont{Di~Sante}},
  \bibinfo{author}{\bibfnamefont{A.}~\bibnamefont{Narayan}},
  \bibinfo{author}{\bibfnamefont{L.}~\bibnamefont{He}},
  \bibinfo{author}{\bibfnamefont{S.}~\bibnamefont{Picozzi}},
  \bibinfo{author}{\bibfnamefont{S.}~\bibnamefont{Sanvito}},
  \bibnamefont{et~al.}, \bibinfo{journal}{NPG Asia Mater}
  \textbf{\bibinfo{volume}{7}}, \bibinfo{pages}{e221} (\bibinfo{year}{2015}).

\bibitem[{\citenamefont{Xiong et~al.}(2016)\citenamefont{Xiong, Kushwaha,
  Krizan, Liang, Cava, and Ong}}]{ong2016}
\bibinfo{author}{\bibfnamefont{J.}~\bibnamefont{Xiong}},
  \bibinfo{author}{\bibfnamefont{S.}~\bibnamefont{Kushwaha}},
  \bibinfo{author}{\bibfnamefont{J.}~\bibnamefont{Krizan}},
  \bibinfo{author}{\bibfnamefont{T.}~\bibnamefont{Liang}},
  \bibinfo{author}{\bibfnamefont{R.~J.} \bibnamefont{Cava}}, \bibnamefont{and}
  \bibinfo{author}{\bibfnamefont{N.~P.} \bibnamefont{Ong}},
  \bibinfo{journal}{EPL (Europhysics Letters)} \textbf{\bibinfo{volume}{114}},
  \bibinfo{pages}{27002} (\bibinfo{year}{2016}).

\bibitem[{\citenamefont{Wang et~al.}(2013)\citenamefont{Wang, Weng, Wu, Dai,
  and Fang}}]{zhizhun2013}
\bibinfo{author}{\bibfnamefont{Z.}~\bibnamefont{Wang}},
  \bibinfo{author}{\bibfnamefont{H.}~\bibnamefont{Weng}},
  \bibinfo{author}{\bibfnamefont{Q.}~\bibnamefont{Wu}},
  \bibinfo{author}{\bibfnamefont{X.}~\bibnamefont{Dai}}, \bibnamefont{and}
  \bibinfo{author}{\bibfnamefont{Z.}~\bibnamefont{Fang}},
  \bibinfo{journal}{Phys. Rev. B} \textbf{\bibinfo{volume}{88}},
  \bibinfo{pages}{125427} (\bibinfo{year}{2013}).

\bibitem[{\citenamefont{Borisenko et~al.}(2014)\citenamefont{Borisenko, Gibson,
  Evtushinsky, Zabolotnyy, B\"uchner, and Cava}}]{borisenko2014}
\bibinfo{author}{\bibfnamefont{S.}~\bibnamefont{Borisenko}},
  \bibinfo{author}{\bibfnamefont{Q.}~\bibnamefont{Gibson}},
  \bibinfo{author}{\bibfnamefont{D.}~\bibnamefont{Evtushinsky}},
  \bibinfo{author}{\bibfnamefont{V.}~\bibnamefont{Zabolotnyy}},
  \bibinfo{author}{\bibfnamefont{B.}~\bibnamefont{B\"uchner}},
  \bibnamefont{and} \bibinfo{author}{\bibfnamefont{R.~J.} \bibnamefont{Cava}},
  \bibinfo{journal}{Phys. Rev. Lett.} \textbf{\bibinfo{volume}{113}},
  \bibinfo{pages}{027603} (\bibinfo{year}{2014}).

\bibitem[{\citenamefont{Neupane et~al.}(2014)\citenamefont{Neupane, Xu, Sankar,
  Alidoust, Bian, Liu, Belopolski, Chang, Jeng, Lin et~al.}}]{neupane2014}
\bibinfo{author}{\bibfnamefont{M.}~\bibnamefont{Neupane}},
  \bibinfo{author}{\bibfnamefont{S.-Y.} \bibnamefont{Xu}},
  \bibinfo{author}{\bibfnamefont{R.}~\bibnamefont{Sankar}},
  \bibinfo{author}{\bibfnamefont{N.}~\bibnamefont{Alidoust}},
  \bibinfo{author}{\bibfnamefont{G.}~\bibnamefont{Bian}},
  \bibinfo{author}{\bibfnamefont{C.}~\bibnamefont{Liu}},
  \bibinfo{author}{\bibfnamefont{I.}~\bibnamefont{Belopolski}},
  \bibinfo{author}{\bibfnamefont{T.-R.} \bibnamefont{Chang}},
  \bibinfo{author}{\bibfnamefont{H.-T.} \bibnamefont{Jeng}},
  \bibinfo{author}{\bibfnamefont{H.}~\bibnamefont{Lin}}, \bibnamefont{et~al.},
  \bibinfo{journal}{Nat Commun} \textbf{\bibinfo{volume}{5}},
  \bibinfo{pages}{4786} (\bibinfo{year}{2014}).

\bibitem[{\citenamefont{Jeon et~al.}(2014)\citenamefont{Jeon, Zhou, Gyenis,
  Feldman, Kimchi, Potter, Gibson, Cava, Vishwanath, and Yazdani}}]{jeon2014}
\bibinfo{author}{\bibfnamefont{S.}~\bibnamefont{Jeon}},
  \bibinfo{author}{\bibfnamefont{B.~B.} \bibnamefont{Zhou}},
  \bibinfo{author}{\bibfnamefont{A.}~\bibnamefont{Gyenis}},
  \bibinfo{author}{\bibfnamefont{B.~E.} \bibnamefont{Feldman}},
  \bibinfo{author}{\bibfnamefont{I.}~\bibnamefont{Kimchi}},
  \bibinfo{author}{\bibfnamefont{A.~C.} \bibnamefont{Potter}},
  \bibinfo{author}{\bibfnamefont{Q.~D.} \bibnamefont{Gibson}},
  \bibinfo{author}{\bibfnamefont{R.~J.} \bibnamefont{Cava}},
  \bibinfo{author}{\bibfnamefont{A.}~\bibnamefont{Vishwanath}},
  \bibnamefont{and} \bibinfo{author}{\bibfnamefont{A.}~\bibnamefont{Yazdani}},
  \bibinfo{journal}{Nat Mater} \textbf{\bibinfo{volume}{13}},
  \bibinfo{pages}{851} (\bibinfo{year}{2014}).

\bibitem[{\citenamefont{Liu et~al.}(2014{\natexlab{a}})\citenamefont{Liu,
  Jiang, Zhou, Wang, Zhang, Weng, Prabhakaran, Mo, Peng, Dudin
  et~al.}}]{liu2014b}
\bibinfo{author}{\bibfnamefont{Z.~K.} \bibnamefont{Liu}},
  \bibinfo{author}{\bibfnamefont{J.}~\bibnamefont{Jiang}},
  \bibinfo{author}{\bibfnamefont{B.}~\bibnamefont{Zhou}},
  \bibinfo{author}{\bibfnamefont{Z.~J.} \bibnamefont{Wang}},
  \bibinfo{author}{\bibfnamefont{Y.}~\bibnamefont{Zhang}},
  \bibinfo{author}{\bibfnamefont{H.~M.} \bibnamefont{Weng}},
  \bibinfo{author}{\bibfnamefont{D.}~\bibnamefont{Prabhakaran}},
  \bibinfo{author}{\bibfnamefont{S.-K.} \bibnamefont{Mo}},
  \bibinfo{author}{\bibfnamefont{H.}~\bibnamefont{Peng}},
  \bibinfo{author}{\bibfnamefont{P.}~\bibnamefont{Dudin}},
  \bibnamefont{et~al.}, \bibinfo{journal}{Nat Mater}
  \textbf{\bibinfo{volume}{13}}, \bibinfo{pages}{677}
  (\bibinfo{year}{2014}{\natexlab{a}}), ISSN \bibinfo{issn}{1476-1122}.

\bibitem[{\citenamefont{Wang et~al.}(2012)\citenamefont{Wang, Sun, Chen,
  Franchini, Xu, Weng, Dai, and Fang}}]{zhizhun2012}
\bibinfo{author}{\bibfnamefont{Z.}~\bibnamefont{Wang}},
  \bibinfo{author}{\bibfnamefont{Y.}~\bibnamefont{Sun}},
  \bibinfo{author}{\bibfnamefont{X.-Q.} \bibnamefont{Chen}},
  \bibinfo{author}{\bibfnamefont{C.}~\bibnamefont{Franchini}},
  \bibinfo{author}{\bibfnamefont{G.}~\bibnamefont{Xu}},
  \bibinfo{author}{\bibfnamefont{H.}~\bibnamefont{Weng}},
  \bibinfo{author}{\bibfnamefont{X.}~\bibnamefont{Dai}}, \bibnamefont{and}
  \bibinfo{author}{\bibfnamefont{Z.}~\bibnamefont{Fang}},
  \bibinfo{journal}{Phys. Rev. B} \textbf{\bibinfo{volume}{85}},
  \bibinfo{pages}{195320} (\bibinfo{year}{2012}).

\bibitem[{\citenamefont{Liu et~al.}(2014{\natexlab{b}})\citenamefont{Liu, Zhou,
  Zhang, Wang, Weng, Prabhakaran, Mo, Shen, Fang, Dai et~al.}}]{yulin2014}
\bibinfo{author}{\bibfnamefont{Z.~K.} \bibnamefont{Liu}},
  \bibinfo{author}{\bibfnamefont{B.}~\bibnamefont{Zhou}},
  \bibinfo{author}{\bibfnamefont{Y.}~\bibnamefont{Zhang}},
  \bibinfo{author}{\bibfnamefont{Z.~J.} \bibnamefont{Wang}},
  \bibinfo{author}{\bibfnamefont{H.~M.} \bibnamefont{Weng}},
  \bibinfo{author}{\bibfnamefont{D.}~\bibnamefont{Prabhakaran}},
  \bibinfo{author}{\bibfnamefont{S.-K.} \bibnamefont{Mo}},
  \bibinfo{author}{\bibfnamefont{Z.~X.} \bibnamefont{Shen}},
  \bibinfo{author}{\bibfnamefont{Z.}~\bibnamefont{Fang}},
  \bibinfo{author}{\bibfnamefont{X.}~\bibnamefont{Dai}}, \bibnamefont{et~al.},
  \bibinfo{journal}{Science} \textbf{\bibinfo{volume}{343}},
  \bibinfo{pages}{864} (\bibinfo{year}{2014}{\natexlab{b}}).

\bibitem[{\citenamefont{Zhang et~al.}(2014)\citenamefont{Zhang, Liu, Zhou, Kim,
  Hussain, Shen, Chen, and Mo}}]{yulin2014b}
\bibinfo{author}{\bibfnamefont{Y.}~\bibnamefont{Zhang}},
  \bibinfo{author}{\bibfnamefont{Z.}~\bibnamefont{Liu}},
  \bibinfo{author}{\bibfnamefont{B.}~\bibnamefont{Zhou}},
  \bibinfo{author}{\bibfnamefont{Y.}~\bibnamefont{Kim}},
  \bibinfo{author}{\bibfnamefont{Z.}~\bibnamefont{Hussain}},
  \bibinfo{author}{\bibfnamefont{Z.-X.} \bibnamefont{Shen}},
  \bibinfo{author}{\bibfnamefont{Y.}~\bibnamefont{Chen}}, \bibnamefont{and}
  \bibinfo{author}{\bibfnamefont{S.-K.} \bibnamefont{Mo}},
  \bibinfo{journal}{Applied Physics Letters} \textbf{\bibinfo{volume}{105}},
  \bibinfo{eid}{031901} (\bibinfo{year}{2014}).

\bibitem[{\citenamefont{Xu et~al.}(2015{\natexlab{a}})\citenamefont{Xu,
  Belopolski, Alidoust, Neupane, Bian, Zhang, Sankar, Chang, Yuan, Lee
  et~al.}}]{hasan2015}
\bibinfo{author}{\bibfnamefont{S.-Y.} \bibnamefont{Xu}},
  \bibinfo{author}{\bibfnamefont{I.}~\bibnamefont{Belopolski}},
  \bibinfo{author}{\bibfnamefont{N.}~\bibnamefont{Alidoust}},
  \bibinfo{author}{\bibfnamefont{M.}~\bibnamefont{Neupane}},
  \bibinfo{author}{\bibfnamefont{G.}~\bibnamefont{Bian}},
  \bibinfo{author}{\bibfnamefont{C.}~\bibnamefont{Zhang}},
  \bibinfo{author}{\bibfnamefont{R.}~\bibnamefont{Sankar}},
  \bibinfo{author}{\bibfnamefont{G.}~\bibnamefont{Chang}},
  \bibinfo{author}{\bibfnamefont{Z.}~\bibnamefont{Yuan}},
  \bibinfo{author}{\bibfnamefont{C.-C.} \bibnamefont{Lee}},
  \bibnamefont{et~al.}, \bibinfo{journal}{Science}
  \textbf{\bibinfo{volume}{349}}, \bibinfo{pages}{613}
  (\bibinfo{year}{2015}{\natexlab{a}}).

\bibitem[{\citenamefont{Lv et~al.}(2015)\citenamefont{Lv, Weng, Fu, Wang, Miao,
  Ma, Richard, Huang, Zhao, Chen et~al.}}]{ding2015}
\bibinfo{author}{\bibfnamefont{B.~Q.} \bibnamefont{Lv}},
  \bibinfo{author}{\bibfnamefont{H.~M.} \bibnamefont{Weng}},
  \bibinfo{author}{\bibfnamefont{B.~B.} \bibnamefont{Fu}},
  \bibinfo{author}{\bibfnamefont{X.~P.} \bibnamefont{Wang}},
  \bibinfo{author}{\bibfnamefont{H.}~\bibnamefont{Miao}},
  \bibinfo{author}{\bibfnamefont{J.}~\bibnamefont{Ma}},
  \bibinfo{author}{\bibfnamefont{P.}~\bibnamefont{Richard}},
  \bibinfo{author}{\bibfnamefont{X.~C.} \bibnamefont{Huang}},
  \bibinfo{author}{\bibfnamefont{L.~X.} \bibnamefont{Zhao}},
  \bibinfo{author}{\bibfnamefont{G.~F.} \bibnamefont{Chen}},
  \bibnamefont{et~al.}, \bibinfo{journal}{Phys. Rev. X}
  \textbf{\bibinfo{volume}{5}}, \bibinfo{pages}{031013} (\bibinfo{year}{2015}).

\bibitem[{\citenamefont{Shekhar et~al.}(2015)\citenamefont{Shekhar, Nayak, Sun,
  Schmidt, Nicklas, Leermakers, Zeitler, Skourski, Wosnitza, Liu
  et~al.}}]{yan2015}
\bibinfo{author}{\bibfnamefont{C.}~\bibnamefont{Shekhar}},
  \bibinfo{author}{\bibfnamefont{A.~K.} \bibnamefont{Nayak}},
  \bibinfo{author}{\bibfnamefont{Y.}~\bibnamefont{Sun}},
  \bibinfo{author}{\bibfnamefont{M.}~\bibnamefont{Schmidt}},
  \bibinfo{author}{\bibfnamefont{M.}~\bibnamefont{Nicklas}},
  \bibinfo{author}{\bibfnamefont{I.}~\bibnamefont{Leermakers}},
  \bibinfo{author}{\bibfnamefont{U.}~\bibnamefont{Zeitler}},
  \bibinfo{author}{\bibfnamefont{Y.}~\bibnamefont{Skourski}},
  \bibinfo{author}{\bibfnamefont{J.}~\bibnamefont{Wosnitza}},
  \bibinfo{author}{\bibfnamefont{Z.}~\bibnamefont{Liu}}, \bibnamefont{et~al.},
  \bibinfo{journal}{Nat Phys} \textbf{\bibinfo{volume}{11}},
  \bibinfo{pages}{645} (\bibinfo{year}{2015}), \bibinfo{note}{letter}.

\bibitem[{\citenamefont{Yang et~al.}(2015)\citenamefont{Yang, Liu, Sun, Peng,
  Yang, Zhang, Zhou, Zhang, Guo, Rahn et~al.}}]{chen2015}
\bibinfo{author}{\bibfnamefont{L.~X.} \bibnamefont{Yang}},
  \bibinfo{author}{\bibfnamefont{Z.~K.} \bibnamefont{Liu}},
  \bibinfo{author}{\bibfnamefont{Y.}~\bibnamefont{Sun}},
  \bibinfo{author}{\bibfnamefont{H.}~\bibnamefont{Peng}},
  \bibinfo{author}{\bibfnamefont{H.~F.} \bibnamefont{Yang}},
  \bibinfo{author}{\bibfnamefont{T.}~\bibnamefont{Zhang}},
  \bibinfo{author}{\bibfnamefont{B.}~\bibnamefont{Zhou}},
  \bibinfo{author}{\bibfnamefont{Y.}~\bibnamefont{Zhang}},
  \bibinfo{author}{\bibfnamefont{Y.~F.} \bibnamefont{Guo}},
  \bibinfo{author}{\bibfnamefont{M.}~\bibnamefont{Rahn}}, \bibnamefont{et~al.},
  \bibinfo{journal}{Nat Phys} \textbf{\bibinfo{volume}{11}},
  \bibinfo{pages}{728} (\bibinfo{year}{2015}), \bibinfo{note}{letter}.

\bibitem[{\citenamefont{Xu et~al.}(2015{\natexlab{b}})\citenamefont{Xu,
  Alidoust, Belopolski, Yuan, Bian, Chang, Zheng, Strocov, Sanchez, Chang
  et~al.}}]{xu2015}
\bibinfo{author}{\bibfnamefont{S.-Y.} \bibnamefont{Xu}},
  \bibinfo{author}{\bibfnamefont{N.}~\bibnamefont{Alidoust}},
  \bibinfo{author}{\bibfnamefont{I.}~\bibnamefont{Belopolski}},
  \bibinfo{author}{\bibfnamefont{Z.}~\bibnamefont{Yuan}},
  \bibinfo{author}{\bibfnamefont{G.}~\bibnamefont{Bian}},
  \bibinfo{author}{\bibfnamefont{T.-R.} \bibnamefont{Chang}},
  \bibinfo{author}{\bibfnamefont{H.}~\bibnamefont{Zheng}},
  \bibinfo{author}{\bibfnamefont{V.~N.} \bibnamefont{Strocov}},
  \bibinfo{author}{\bibfnamefont{D.~S.} \bibnamefont{Sanchez}},
  \bibinfo{author}{\bibfnamefont{G.}~\bibnamefont{Chang}},
  \bibnamefont{et~al.}, \bibinfo{journal}{Nat Phys}
  \textbf{\bibinfo{volume}{11}}, \bibinfo{pages}{748}
  (\bibinfo{year}{2015}{\natexlab{b}}), \bibinfo{note}{article}.

\bibitem[{\citenamefont{Li et~al.}(2016{\natexlab{b}})\citenamefont{Li, He, Lu,
  Zhang, Liu, Ma, Fan, Shen, and Wang}}]{hui2016}
\bibinfo{author}{\bibfnamefont{H.}~\bibnamefont{Li}},
  \bibinfo{author}{\bibfnamefont{H.}~\bibnamefont{He}},
  \bibinfo{author}{\bibfnamefont{H.-Z.} \bibnamefont{Lu}},
  \bibinfo{author}{\bibfnamefont{H.}~\bibnamefont{Zhang}},
  \bibinfo{author}{\bibfnamefont{H.}~\bibnamefont{Liu}},
  \bibinfo{author}{\bibfnamefont{R.}~\bibnamefont{Ma}},
  \bibinfo{author}{\bibfnamefont{Z.}~\bibnamefont{Fan}},
  \bibinfo{author}{\bibfnamefont{S.-Q.} \bibnamefont{Shen}}, \bibnamefont{and}
  \bibinfo{author}{\bibfnamefont{J.}~\bibnamefont{Wang}}, \bibinfo{journal}{Nat
  Commun} \textbf{\bibinfo{volume}{7}} (\bibinfo{year}{2016}{\natexlab{b}}).

\bibitem[{\citenamefont{Li et~al.}(2015)\citenamefont{Li, Wang, Liu, Wang,
  Liao, and Yu}}]{li2015}
\bibinfo{author}{\bibfnamefont{C.-Z.} \bibnamefont{Li}},
  \bibinfo{author}{\bibfnamefont{L.-X.} \bibnamefont{Wang}},
  \bibinfo{author}{\bibfnamefont{H.}~\bibnamefont{Liu}},
  \bibinfo{author}{\bibfnamefont{J.}~\bibnamefont{Wang}},
  \bibinfo{author}{\bibfnamefont{Z.-M.} \bibnamefont{Liao}}, \bibnamefont{and}
  \bibinfo{author}{\bibfnamefont{D.-P.} \bibnamefont{Yu}},
  \bibinfo{journal}{Nat Commun} \textbf{\bibinfo{volume}{6}},
  \bibinfo{pages}{10137} (\bibinfo{year}{2015}).

\bibitem[{\citenamefont{Wang et~al.}(2016)\citenamefont{Wang, Li, Yu, and
  Liao}}]{wang2016}
\bibinfo{author}{\bibfnamefont{L.-X.} \bibnamefont{Wang}},
  \bibinfo{author}{\bibfnamefont{C.-Z.} \bibnamefont{Li}},
  \bibinfo{author}{\bibfnamefont{D.-P.} \bibnamefont{Yu}}, \bibnamefont{and}
  \bibinfo{author}{\bibfnamefont{Z.-M.} \bibnamefont{Liao}},
  \bibinfo{journal}{Nat Commun} \textbf{\bibinfo{volume}{7}},
  \bibinfo{pages}{10769} (\bibinfo{year}{2016}).

\bibitem[{\citenamefont{Cano et~al.}(unpublished)\citenamefont{Cano, Bradlyn,
  Wang, Hirschberger, Ong, and Bernevig}}]{cano16}
\bibinfo{author}{\bibfnamefont{J.}~\bibnamefont{Cano}},
  \bibinfo{author}{\bibfnamefont{B.}~\bibnamefont{Bradlyn}},
  \bibinfo{author}{\bibfnamefont{Z.}~\bibnamefont{Wang}},
  \bibinfo{author}{\bibfnamefont{M.}~\bibnamefont{Hirschberger}},
  \bibinfo{author}{\bibfnamefont{N.}~\bibnamefont{Ong}}, \bibnamefont{and}
  \bibinfo{author}{\bibfnamefont{B.}~\bibnamefont{Bernevig}},
  \bibinfo{journal}{arXiv:1604.08601}  (\bibinfo{year}{unpublished}).

\end{thebibliography}

%%%%%%%%%%%%%%%%%%%%%%%%%%%%%%%%%%%%%%%%%%%%%%%%%%
\appendix
\section{Model parameters}

\begin{figure*}[t]
\includegraphics[width = 14.5cm]{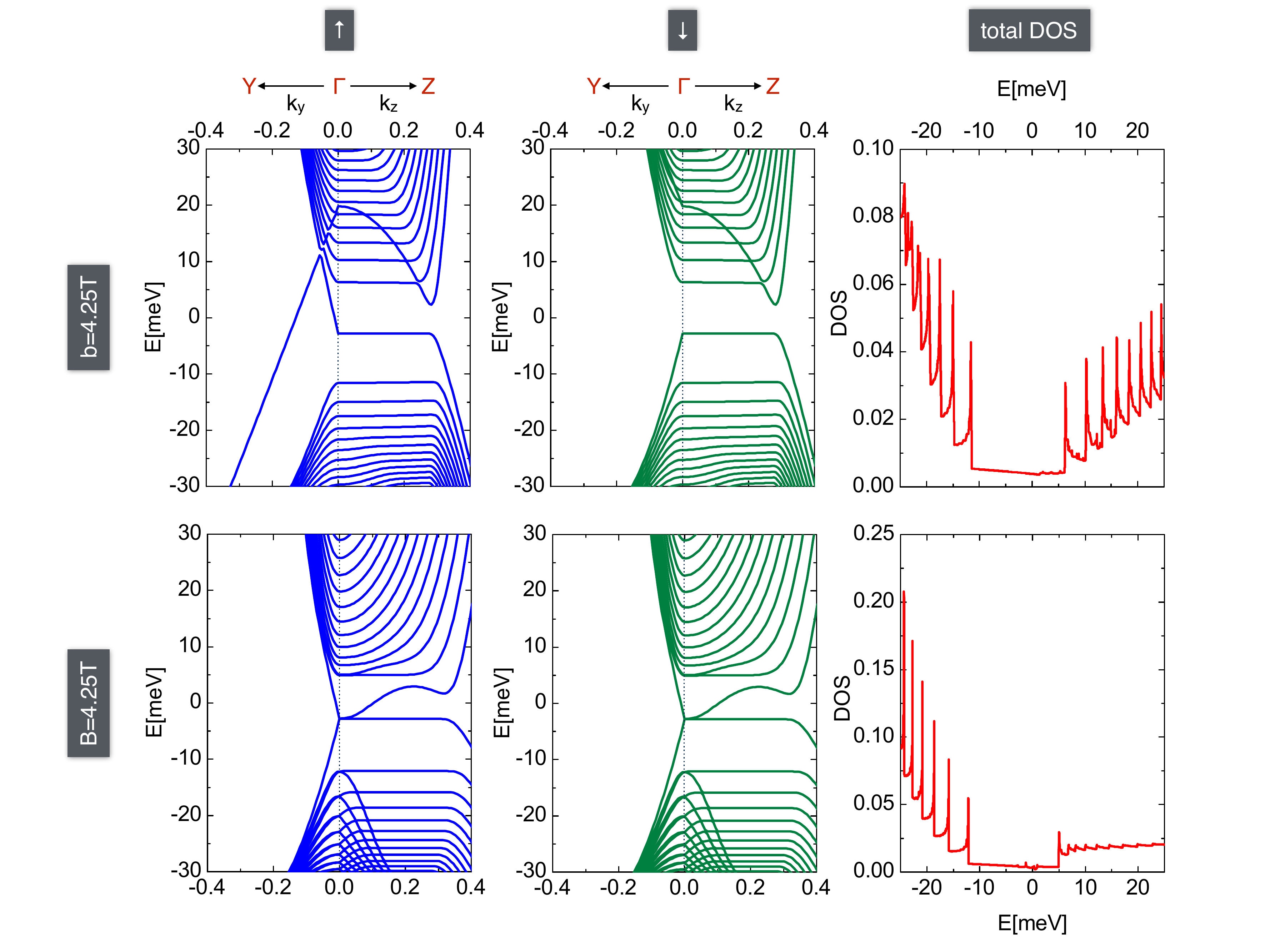}
\caption{Bandstructure and density of states for the model of
  Cd$_3$As$_2$ with the particle-hole asymmetric part
  \eqref{eq:ph_asymmetry} included. Top row is for the pseudomagnetic
  field $b=4.25$T (corresponding to $\alpha=0.04$, stronger strain than in
  the main text), and bottom row is for real magnetic field $B=4.25$T. From left to right -- bandstructure for spin up band, bandstructure for spin down band, and normalized total DOS.
}\label{fig5}
\end{figure*} 

We model Cd$_3$As$_2$ using the Hamiltonian \eqref{h1} with parameters
taken from the first principles band structure calculations
\cite{cano16,pikulin2016} and with the lattice constants corresponding
to the actual material, $a_{x, y} = 3$\AA, $a_z=5$\AA. This implies
that the constants used in \eqref{h3} are: $\Lambda = 0.296$eV, $t_0 = - 7.4811$eV, $t_1 = 1.5016$eV, and $t_2 = 3$eV. Additionally, we model the particle-hole asymmetry of the real Cd$_3$As$_2$ using
\begin{align}
\epsilon_\bk = r_0 + r_1 \cos a_z k_z + r_2(\cos a_x k_x + \cos a_y k_y),\label{eq:ph_asymmetry}
\end{align}
with $r_0=5.9439$eV, $r_1=-0.8472$eV, and $r_2=-2.5556$eV. 

The results
for the dispersion and DOS for the realistic particle-hole asymmetric
case are shown in Fig. \ref{fig5}. We note the similarity of the
results with those displayed in the main text and in
Fig. \ref{fig5}. Specifically, both the real magnetic field $B$ and
the strain-induced pseudomagnetic field $b$ give rise to pronounced
Landau levels. We thus conclude that all our predictions remain valid for this realistic Dirac semimetal.

\section{Conductivity calculation}

First we obtain the analytical expression for the conductivity of
Dirac-Landau levels in the bulk and in the continuum limit. From the dispersion relation \eqref{lan1} we obtain the velocity in $y$ direction,
\begin{align}
v_n^{s}(k_y) = \frac{1}{\hbar} \frac{\partial E_n^s}{\partial k_y} = s v_y \frac{k_y}{\sqrt{k_y^2 + n \Omega}},
\end{align}
where $s$ is the sign of the energy and $\Omega = \frac{2 e B}{\hbar c} \frac{v_x v_z}{v_y^2}$.
Then we use the familiar formula for the DC conductivity $\sigma_{yy}$
due to the $n$'th Landau level in the relaxation time approximation
\begin{align}
\sigma_n^s(\mu) = e^2 \int \frac{dk_y}{2\pi} \tau_n^s(E_n^2(k_y)) (v_n^s(k_y))^2 \left(- \frac{\partial f(E - \mu)}{\partial E}\right)_{E_n^s(k_y)}.\label{eq:sigma}
\end{align}
Here $f(\epsilon)$ is the Fermi function. We assume zero temperature,
angle-independent relaxation time, and substitute the dispersion
relation to obtain
\begin{align}
\sigma_n^s(\mu) = e^2 \tau_n^s \int \frac{dk_y}{2\pi} \frac{v_y^2 k_y^2}{k_y^2 + n \Omega} \delta(E_n^s(k_y) - \mu).
\end{align}
After the change of the integration variable from $k_y$ to $E_n^s(k_y)$ and integration we find
\begin{align}
\sigma_n^s(\mu) = \frac{e^2 \tau_n^s v_y}{\hbar \pi} \left.\frac{\mathrm{Re}\sqrt{E_n^2 - n \hbar^2 v_y^2 \Omega}}{E_n^s}\right|_{E_n^s=\mu},
\end{align}
and the total conductivity is
\begin{align}
\sigma = \frac{2e^2 v_y}{h}\sum_n \tau_n(\mu) \mathrm{Re}\sqrt{\frac{\mu^2 - n \hbar^2 v_y^2 \Omega}{\mu^2}}.
\end{align}
Finally, we estimate the relaxation time in the lowest order Born approximation
\begin{align}
\frac{1}{\tau} = 2\pi D(\mu) n_{\rm imp} C,
\end{align}
where $D(\mu)$ is the density of states at the Fermi level and $n_{\rm
  imp}$ is the impurity concentration. Constant $C$ depends on the details of scattering from impurities. Thus the final formula we use for the conductivity computation in Fig. \ref{fig3} is
\begin{align}
\sigma_{yy} = \frac{e^2 v_y}{\pi h D(\mu) n_{\rm imp} C}\sum_n\mathrm{Re}\sqrt{\frac{\mu^2 - n\hbar^2 v_y^2 \Omega}{\mu^2}}.
\end{align}

Numerically we use the same formula \eqref{eq:sigma}, but input the actual velocities and energies into it.

\section{Quantum oscillations above Lifshitz transition}

\begin{figure*}[t]
\includegraphics[width = 0.8\linewidth]{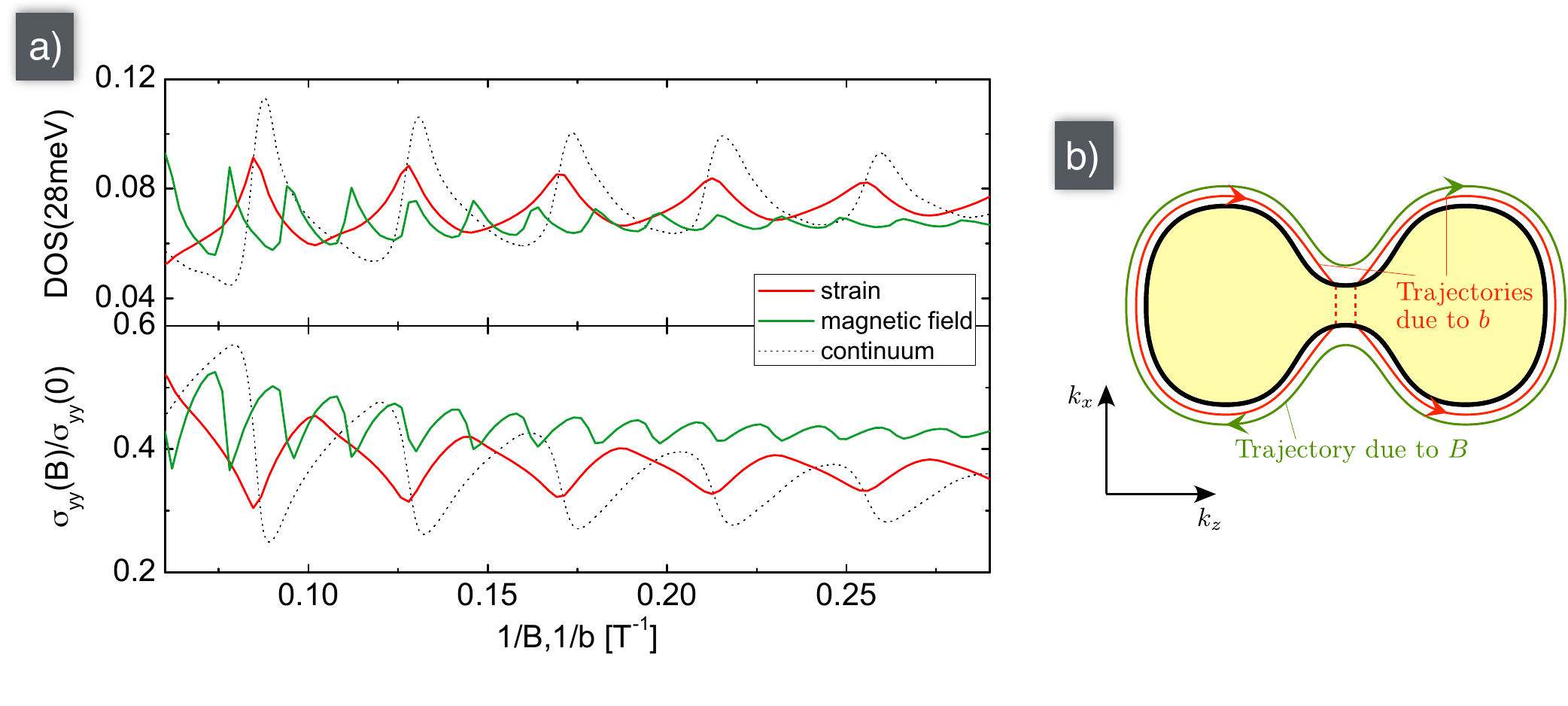}
\caption{a) QO above the Lifshits transition due to ordinary magnetic field and due to the gauge field. Period difference by more than a factor of 2 is seen. The low-energy analytics does not apply anymore, as expected. b) Corresponding hypothesized quasiclassical trajectories of electrons in the Brillouin zone. Green -- for $B_y$ field, and red -- for $b_y$ field.
}\label{fig6}
\end{figure*}

In this section we present the results for QO at energy $28$meV, above the Lifshitz transition (at approximatly $20$meV). In Fig. \ref{fig6}a we see that the area of the Fermi surface causing the oscillations in $B$ and $b$ fields is different by slightly larger than a factor of 2. For the external magnetic field case the effective area of the Fermi surface is approximately doubled as compared to the gauge field. Strain couples only to the linear part of the Hamiltonian as a gauge field, therefore only the oscillations around each of the Weyl points are possible. Notice also that the electron in the pseudmagnetic field  travels clockwise around one of the Weyl points and counterclockwise around the another. The precise nature of the corresponding quasiclassical trajectories above the Lifshitz transition is therefore an interesting open question which we leave for further study. We speculate that they include tunneling between the opposite points of the Fermi surface as depicted in Fig. \ref{fig6}b. Such trajectories would define an extremal area consistent with our numerical results.
% Corresponding quasiclassical trajectories of the electrons are depicted in Fig. \ref{fig6}b. Notice that the electron in a gauge field travels clockwise around one of the Weyl points and counterclockwise around another. The trajectory includes tunneling between the opposite points of Fermi surface.
\newline
\section{Equivalence of external and gauge fields}

\begin{figure}[t]
\includegraphics[width = 0.8\linewidth]{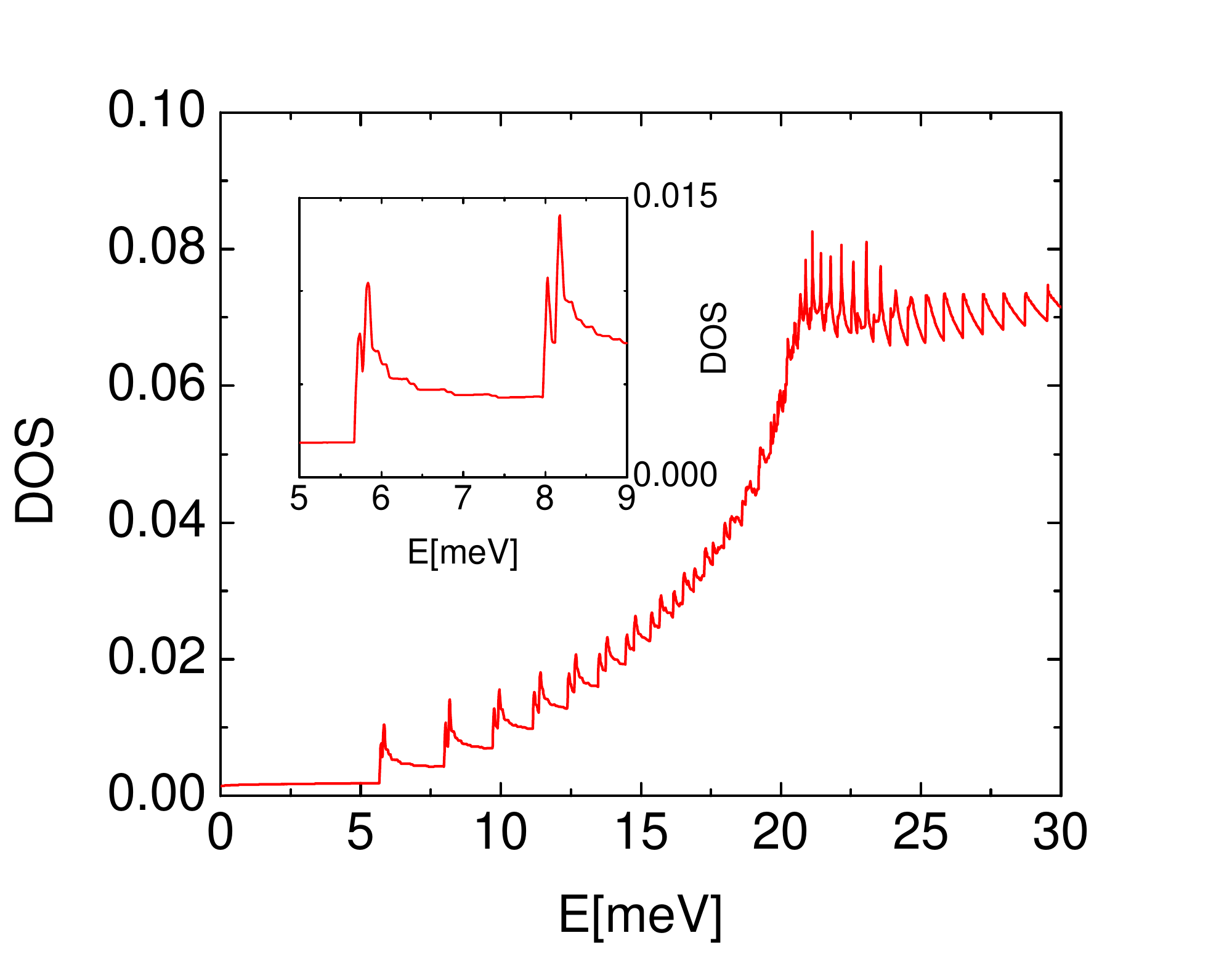}
\caption{Normalized density of states for both fields present, $B=1$T and $b=0.0184$T. Each of the DOS peaks due to ordinary magnetic field splits due to torsion thus proving the equivalence of the external and gauge fields. Inset gives closer view of the first two peaks.
}\label{fig7}
\end{figure}

In this section we additionally substantiate the proposed equivalence of $b$ and $B$ fields and suggest an additional experimental test. We propose to apply external magnetic field of fixed strength and then slowly turn on strain (or vice versa, whichever is more convenient in a particular experimental design). This will result in splitting of the first peak in DOS as seen in Fig. \ref{fig7}. This happens because the two Weyl cones will feel different effective magnetic fields, $B+b$ and $B-b$, which result in two independent sequences of peaks in DOS. Observation of the splitting would prove the identical nature of the gauge and external magnetic fields in each of the Weyl cones, and establish that the two cones feel opposite effective field due to $b$. 

%Additionally, such measurement involving both the fields would provide a natural way to relate strain to the effective magnetic field $b$. Importantly, such method  does not require knowledge of the constants in \eqref{lan1}. We suggest to crank up the gauge field until the two peaks, one moving up from the lowest initial peak, and one moving down from the second initial peak, merge. For their energies to coincide, according to \eqref{lan1}, one needs the following condition to be satisfied:
%\begin{align}
%\sqrt{2|B-b|} = \sqrt{|B+b|},
%\end{align}
%therefore giving $b=B/3$. This would create a correspondence between the applied stress and the generated gauge field which can be used for further characterization of the device.

\end{document}